\documentclass[12pt,a4paper]{article}

\usepackage{amsmath,amsthm,amssymb,amscd,a4wide}
\usepackage[dvips]{graphicx}

\usepackage{dsfont}
\usepackage{mathrsfs}

\DeclareMathOperator{\im}{Im}
\DeclareMathOperator{\re}{Re}
\DeclareMathOperator{\tr}{tr}

\newcommand{\bm}{\boldsymbol}
\newcommand{\openone}{\mathds{1}}

\numberwithin{equation}{section}

\begin{document}

\thispagestyle{empty}
\noindent
ULM-TP/05-3\\
September 2005\\

\vspace*{1cm}

\begin{center}

{\LARGE\bf   Stability of wave packet dynamics  \\ 
\vspace*{3mm} under perturbations} \\
\vspace*{3cm}
{\large Jens Bolte}%
\footnote{E-mail address: {\tt jens.bolte@uni-ulm.de}}
{\large and Tobias Schwaibold}%
\footnote{E-mail address: {\tt tobias.schwaibold@uni-ulm.de}}

\vspace*{1cm}

Abteilung Theoretische Physik\\
Universit\"at Ulm, Albert-Einstein-Allee 11\\
D-89069 Ulm, Germany 
\end{center}

\vfill

\begin{abstract}
We introduce a novel method to investigate the stability of wave packet 
dynamics under perturbations of the Hamiltonian. Our approach relies on
semiclassical approximations, but is non-perturbative. Two separate 
contributions to the quantum fidelity are identified: one factor derives 
from the dispersion of the wave packets, whereas the other factor is
determined by the separation of a trajectory of the perturbed classical 
system away from a corresponding unperturbed trajectory. We furthermore 
estimate both contributions in terms of classical Lyapunov exponents
and find a decay of fidelity that is, generically, at least exponential, 
but may also be doubly exponential. The latter case is shown to be realised 
for inverted harmonic oscillators.
\end{abstract}

\newpage

\section{Introduction}
\label{intro}
It has long been appreciated that, in contrast to chaotic classical 
dynamics, the time evolution of a quantum system shows no sensible
dependence on initial conditions. This follows immediately from the 
unitarity of quantum dynamics. Hence, rather than being concerned with 
instabilities at large times, the notion of \textit{quantum chaos} is 
commonly reserved for semiclassical studies that aim at relating statistical
properties of stationary quantum states to dynamical properties of
chaotic classical systems (see, e.g., \cite{Sto99,Haa01}). More recently,
however, the behaviour of quantum time evolutions at large times has
attracted an increasing attention (see, e.g., 
\cite{ComRob97,BouRob02,BolGla05,Sch05}). Both in the dynamics of 
observables (Heisenberg picture) and in the evolution of wave functions 
(Schr\"odinger picture) it has been proven that there exists a time scale 
(the so-called \textit{Ehrenfest time}), depending on the semiclassical 
parameter $\hbar$, below which the quantum dynamics can be well approximated 
in terms of the associated classical time evolution. Moreover, if the 
classical dynamics are chaotic, this time scale is inversely proportional 
to a suitable classical Lyapunov exponent.

Some time ago Peres suggested \cite{Per84} that instead of studying the 
behaviour of quantum dynamics under a change of initial conditions one 
should investigate its stability under perturbations of the Hamiltonian: 
Suppose that an initial state $\psi$ is evolved under the unitary dynamics 
$\hat U_0(t)$ generated by the quantum Hamiltonian $\hat H_0$, one compares 
this with the evolution $\hat U_{\varepsilon}(t)\psi$ determined by the 
perturbed Hamiltonian $\hat H_{\varepsilon}= \hat H_0 +\varepsilon\hat V$. 
Here $\hat V$ is a perturbation of unit strength, and $\varepsilon$ is a 
variable parameter. Then the \textit{quantum fidelity}
\begin{equation}
\label{fidelity}
 F(t) = |\langle \psi,\hat U_{\varepsilon}(t)^{-1}\hat U_0 (t)\psi\rangle|^2
\end{equation}
measures how sensibly the dynamics reacts to this perturbation. It can also 
be viewed as a means to quantify to what extent the initial state can be
recovered after it has been propagated for a time $t$ with the unperturbed
dynamics, and then the time evolution is reversed with a perturbation  
being turned on. For that reason the quantity (\ref{fidelity}) is also 
known as \textit{quantum Loschmidt echo}. 

Peres analysed $F(t)$ in perturbation theory, and found an initial decay
$F(t)\sim 1-C_{\hat V,\psi}(\varepsilon t/\hbar)^2$. Since the reliability 
of perturbative results requires $\varepsilon t/\hbar$ to be small, one 
could view Peres' result as indicating a Gaussian decay of the fidelity on 
an initial time scale that depends on $\varepsilon$ and $\hbar$. Later work 
focused on time scales beyond this perturbative regime or on strong 
perturbations, respectively, and found an exponential decay
\cite{JacSilBee01,JalPas01,CerTom02,CucLewMucPasVal02}. Its rate is either 
determined by Fermi's golden rule \cite{JacSilBee01} or, for stronger 
perturbations, by a classical Lyapunov exponent 
\cite{JalPas01,CucLewMucPasVal02}. Further studies related the behaviour
of the fidelity to the decay of (quantum as well as classical) correlations 
\cite{ProZni02}. All of these results rest on a number of approximations and 
assumptions. Hence the precise time scales for the different regimes 
depend on various factors as, e.g., initial states, strength of perturbation, 
averages over random perturbations, and dynamical properties of the 
corresponding classical dynamics. 

Here our principal aim is to develop an alternative approach to the decay 
of quantum fidelity for particular initial states. The method that we shall 
introduce below is non-perturbative (quantum mechanically as well as 
classically) and employs only semiclassical approximations with a rigorous 
control over the errors. Previous semiclassical studies of fidelity decay 
used the Van~Vleck-Gutzwiller propagator for the time evolution of Gaussian 
initial states \cite{JalPas01,CerTom02,CucLewMucPasVal02,VanHel03}. This 
procedure takes care of the leading term in an expansion in powers of $\hbar$, 
with an error that is, formally, smaller by a factor of $\hbar$. For finite 
times this is indeed true, but there is no analytical control over the 
semiclassical error that arises in estimates of the fidelity decay at large 
times. Our method, however, allows to bound the semiclassical error in terms 
of the linear stability of an associated classical dynamics. It can in
particular be applied to Gaussian states. And although our approach requires
no particular assumptions about the nature of the classical dynamics, we are 
mostly interested in the case of chaotic (i.e., exponentially unstable)
classical trajectories. In that case we find a decay of fidelity prior to 
the Ehrenfest time that generically is at least exponential. But it may 
also be doubly exponential, which we show to be the case in the example of 
inverted harmonic oscillators.

This paper is organised as follows: In Section~\ref{sec1} we introduce the 
wave packets that we shall consider as initial states and review their 
semiclassical dynamics. The decay of quantum fidelity is investigated in 
Section~\ref{sec2}, with an emphasis on the behaviour of Gaussian states. 
An exact calculation of the fidelity for inverted harmonic oscillators is 
performed in Section~\ref{sec3}. We finally summarise our findings in
Section~\ref{sec4}. Three appendices are devoted to a number of technical
considerations: We first review the metaplectic representation, then discuss 
a transformation of positive-definite, symmetric matrices to a diagonal form,
and finally collect estimates of matrix norms and singular values.

\section{Localised wave packets}
\label{sec1}
The initial states to which our approach applies are wave packets with a
localisation both in position and momentum. By this we understand a 
concentration of the quantum state in a suitable phase space representation 
on a single point, when the semiclassical limit is performed by passing to 
a small (effective) Planck's constant $\hbar$. We specify the wave packets 
in terms of normalised, smooth, and rapidly decreasing functions 
$\phi(\bm{x})$ (Schwartz test functions) of $\bm{x}\in\mathbb{R}^d$.
Examples for this are provided by the Gaussian functions 
\begin{equation}
\label{Gauss}
 \phi^{Z}(\bm{x})= \Bigl(\frac{\det\im Z}{\pi^d}\Bigr)^{1/4}\,
                   e^{\frac{i}{2}\bm{x}\cdot Z\bm{x}} \ ,
\end{equation}
where $Z$ is a complex, symmetric $d\times d$ matrix with positive-definite 
imaginary part.

For the purpose of semiclassical investigations we introduce the scaling 
\begin{equation}
\label{scale}
 \phi^{\hbar}(\bm{x}) = \hbar^{-d/4}\phi\bigl(\bm{x}/\sqrt{\hbar}\bigr) \ .
\end{equation}
This produces quantum states that are semiclassically concentrated at zero 
in position and in momentum. Such a phase space localisation is best analysed 
in the Wigner representation 
\begin{equation}
\label{Wigner}
 W[\phi^{\hbar}] (\bm{\xi},\bm{x}) = 
 \int\overline{\phi^{\hbar}(\bm{x}-\tfrac{\bm{y}}{2})}\,\phi^{\hbar}
 (\bm{x}+\tfrac{\bm{y}}{2})\,e^{-\frac{i}{\hbar}\bm{y}\cdot\bm{\xi}}\ dy 
\end{equation}
which, if multiplied by $(2\pi\hbar)^{-d}$, converges to 
$\delta(\bm{\xi},\bm{x})$ as $\hbar\to 0$. A subsequent application of the 
phase space translation 
\begin{equation}
\label{phasetrans}
 \hat D(\bm{p},\bm{q}) =  
 e^{-\frac{i}{\hbar}(\bm{q}\cdot\hat{\bm{P}}-\bm{p}\cdot\hat{\bm{Q}})}
\end{equation}
therefore yields a wave packet 
\begin{equation}
\label{PStranslates}
\begin{split}
 \phi^\hbar_{(\bm{p},\bm{q})}(\bm{x}) 
 &= e^{-\frac{i}{2\hbar}\bm{p}\cdot\bm{q}}\,
    \hat D(\bm{p},\bm{q})\,\phi^{\hbar}(\bm{x}) \\
 &= e^{\frac{i}{\hbar}\bm{p}\cdot(\bm{x}-\bm{q})}\,\phi^{\hbar}
    (\bm{x}-\bm{q})  
\end{split}
\end{equation}
with phase space localisation at the point $(\bm{p},\bm{q})$. The phase 
convention made here is introduced for convenience; it merely simplifies 
some of the expressions below. 

The time evolution of such a state, generated by a quantum Hamiltonian 
$\hat H$ that arises as a Weyl quantisation of a classical Hamiltonian 
$H(\bm{p},\bm{q})$, 
\begin{equation}
\label{Weyl}
 \hat H\psi (\bm{x}) = \iint H\bigl(\bm{p},\tfrac{\bm{x}+\bm{y}}{2}
 \bigr)\,e^{\frac{i}{\hbar}\bm{p}\cdot(\bm{x}-\bm{y})}\,\psi(\bm{y})\ 
 \frac{dp\,dy}{(2\pi\hbar)^d}\ ,
\end{equation}
can be determined semiclassically \cite{PauUri96,ComRob97} to be
\begin{equation}
\label{scltimeevol}
 \hat U(t)\phi^\hbar_{(\bm{p},\bm{q})} =
 e^{\frac{i}{\hbar}R_t}\hat D(\bm{p}_t,\bm{q}_t)\,\hat\mu(S_t)\,\phi^\hbar 
 + O_t(\sqrt{\hbar}) \ .
\end{equation}
The main term on the r.h.s. is again a wave packet of the type 
(\ref{PStranslates}), but now localised at $(\bm{p}_t,\bm{q}_t)$. This is 
the point on the trajectory emerging in time $t$ from the initial point 
$(\bm{p},\bm{q})$ under the classical dynamics generated by the Hamiltonian 
$H(\bm{p},\bm{q})$. Moreover, 
\begin{equation}
\label{action}
 R_t = \int_0^t \bigl( \dot{\bm{q}}_s\cdot\bm{p}_s - H(\bm{p}_s ,\bm{q}_s)
       \bigr)\ ds 
\end{equation}
is the action of this trajectory and $S_t$ is the associated stability matrix. 
This is a real, symplectic $2d\times 2d$ matrix that arises as a solution of
\begin{equation}
\label{Sdef}
 \dot{S}_t = J\,H''(\bm{p}_t ,\bm{q}_t)\, S_t \ ,\quad S_0 = \openone\ .
\end{equation}
Here $J$ is the symplectic unit (\ref{Jdef}) and $H''(\bm{p},\bm{q})$ is 
the Hessian matrix of the Hamiltonian. Equivalently, the stability matrix is 
given as
\begin{equation}
\label{Salt}
 S_t = \begin{pmatrix} \frac{\partial\bm{p}_t}{\partial\bm{p}} & 
       \frac{\partial\bm{p}_t}{\partial\bm{q}} \\
       \frac{\partial\bm{q}_t}{\partial\bm{p}} & 
       \frac{\partial\bm{q}_t}{\partial\bm{q}} \end{pmatrix} \ .
\end{equation}

The wave packet at time $t$ on the r.h.s. of (\ref{scltimeevol}) arises from 
the initial state $\phi^\hbar$ through the application of a unitary operator 
consisting of two contributions: the first factor is the metaplectic operator 
$\hat\mu(S_t)$ that provides a double valued representation of the symplectic 
group (of linear canonical transformations), see Appendix \ref{appa} and 
\cite{Lit86,Fol89,ComRob05b} for details. As can be drawn from
(\ref{Wignermeta}), a metaplectic operator does not change the semiclassical 
phase space localisation. In (\ref{scltimeevol}) it is rather responsible 
for the dispersion of the wave packet. The second factor, 
$\hat D(\bm{p}_t,\bm{q}_t)$, then provides a translation of the wave packet 
along the classical trajectory. Finally, the error term $O_t(\sqrt{\hbar})$ 
stands for a vector whose norm can be estimated from above by 
$K(t)\,\sqrt{\hbar}$. The function $K(t)>0$ contains the linear stability 
of the trajectory $(\bm{p}_t,\bm{q}_t)$. If the latter is exponentially 
unstable with maximal Lyapunov exponent $\lambda>0$, the function $K(t)$ 
grows like $t\,e^{3\lambda t}$ as $t\to\infty$ \cite{ComRob97}. Therefore, 
as long as $t\ll T_E(\hbar)$, with an Ehrenfest time 
$T_E(\hbar)=|\log\hbar|/6\lambda$, the error term remains small. In
contrast, if the trajectory is stable (in an integrable system or on a 
KAM-torus) the growth of $K(t)$ is algebraic (like $t^4$) and hence
$T_E(\hbar)=C\,\hbar^{-1/8}$. In any case, this finding enables one to 
extend the validity of the semiclassical evolution (\ref{scltimeevol}) to 
infinite times, when $\hbar\to 0$. We remark that the main term in 
(\ref{scltimeevol}) actually is the leading contribution in a systematic 
semiclassical expansion \cite{ComRob97}. If this is carried on to the $N$-th 
term, the error is $O_t(\hbar^{N/2})$ and can also be controlled up to 
$T_E(\hbar)$.   
\section{Fidelity decay}
\label{sec2}
The quantum fidelity of an initial wave packet of the type (\ref{PStranslates})
can most conveniently be calculated in the Wigner representation,
\begin{equation}
\label{fidelitycalc1}
 F(t) = |\langle \hat U_0 (t)\phi^\hbar_{(\bm{p},\bm{q})},
         \hat U_{\varepsilon}(t)\phi^\hbar_{(\bm{p},\bm{q})}\rangle|^2 
      = \iint W[\hat U_0(t)\phi^\hbar_{(\bm{p},\bm{q})}](\bm{\xi},\bm{x})\,
         W[\hat U_{\varepsilon}(t)\phi^\hbar_{(\bm{p},\bm{q})}]
         (\bm{\xi},\bm{x})\ \frac{d\xi\,dx}{(2\pi\hbar)^d}\ . 
\end{equation}
We now introduce the semiclassical result (\ref{scltimeevol}) for the 
perturbed and for the unperturbed time evolution, respectively, to 
this expression. The corresponding unperturbed and perturbed classical 
trajectories shall be denoted as $(\bm{p}^0_t,\bm{q}^0_t)$ and 
$(\bm{p}^{\varepsilon}_t,\bm{q}^{\varepsilon}_t)$. Exploiting the behaviour
\begin{equation}
\label{wignertrans}
 W[\hat D(\bm{p},\bm{q})\psi](\bm{\xi},\bm{x}) = 
 W[\psi](\bm{\xi}-\bm{p},\bm{x}-\bm{q})
\end{equation}
of a quantum state $\psi$ in the Wigner representation under phase space
translations, one may change variables and define $(\delta\bm{p}_t,
\delta\bm{q}_t)=(\bm{p}_t^0 - \bm{p}_t^{\varepsilon},\bm{q}_t^0 - 
\bm{q}_t^{\varepsilon})$. For the leading semiclassical contribution one
thus obtains
\begin{equation}
\label{fidelitycalc2}
 F_{\mathrm{scl}}(t) = \iint W[\hat\mu(S^0_t)\phi^\hbar]
 (\bm{\xi}-\delta\bm{p}_t,\bm{x}-\delta\bm{q}_t)\,
 W[\hat\mu(S^{\varepsilon}_t)\phi^\hbar](\bm{\xi},\bm{x})\ 
 \frac{d\xi\,dx}{(2\pi\hbar)^d} \ .
\end{equation}
Since semiclassically the Wigner representations of localised wave packets 
approach $\delta$-functions, after a division by $(2\pi\hbar)^d$ the result 
(\ref{fidelitycalc2}) can be viewed as a smeared out classical fidelity that 
measures the separation $(\delta\bm{p}_t,\delta\bm{q}_t)$ of the perturbed 
trajectory from the unperturbed one. 

For a more detailed study of the expression (\ref{fidelitycalc2}) we now
restrict ourselves to Gaussian initial states of the form (\ref{Gauss}) with 
the scaling (\ref{scale}). The action of a metaplectic operator on such 
states can be calculated explicitly \cite{Lit86,Fol89,ComRob05b},
\begin{equation}
\label{metaplGauss}
 \hat\mu(S)\phi^{Z,\hbar} = e^{i\frac{\pi}{2}\sigma}\,\phi^{S[Z],\hbar} \ .
\end{equation}
Here $S[Z]$ denotes a map, given by the symplectic matrix $S$, on the space 
of complex, symmetric matrices with positive-definite imaginary part to 
itself,
\begin{equation}
\label{Siegelop}
 S[Z] = (AZ+B)(CZ+D)^{-1}\ ,\quad S =\begin{pmatrix}A&B\\C&D\end{pmatrix}\ .
\end{equation}
Furhermore, $\sigma$ is a Maslov phase defined through 
\begin{equation}
\label{Maslov}
e^{i\frac{\pi}{2}\sigma}=\Bigl(\frac{\det\im Z}{\det\im S[Z]}\Bigl)^{1/4}
\,\bigl(\det(CZ+D)\bigr)^{-1/2} \ .
\end{equation}
The Wigner transform of such a Gaussian state is well known to be a Gaussian 
in phase space,
\begin{equation}
\label{WignerGauss}
 W[\phi^{Z,\hbar}](\bm{\xi},\bm{x}) = 2^d e^{-\frac{1}{\hbar}
 (\bm{\xi},\bm{x})\cdot G_Z(\bm{\xi},\bm{x})} \ ,
\end{equation}
where
\begin{equation}
\label{WignerGaussG}
 G_Z = \begin{pmatrix} 
      (\im Z)^{-1} & -(\im Z)^{-1}\re Z\\
       -\re Z(\im Z)^{-1}& \im Z +\re Z(\im Z)^{-1}\re Z
       \end{pmatrix} 
\end{equation}
is a symmetric, symplectic, and positive-definite $2d\times 2d$ matrix with
unit determinant. The behaviour of (\ref{WignerGaussG}) under the 
transformation (\ref{Siegelop}) can be inferred from an application of a
metaplectic operator to a Gaussian state in the Wigner representation
(\ref{WignerGauss}). This way, from (\ref{metaplGauss}) and (\ref{Wignermeta}) 
one concludes that
\begin{equation}
\label{Gtrans}
 G_{S[Z]}= (S^{-1})^T\,G_Z\,S^{-1} \ .
\end{equation}

Now, (\ref{fidelitycalc2}) is a Gaussian integral that can immediately be 
evaluated, and the result may be factorised according to 
\begin{equation}
\label{factor}
 F_{\mathrm{scl}}(t) = F_{\mathrm{disp}}(t)\,F_{\mathrm{class}}(t) \ .
\end{equation}
The first factor 
\begin{equation}
\label{F_disp}
 F_{\mathrm{disp}}(t) =  \Bigl(\det\bigl( G_{S^0_t [Z]}+
 G_{S^\varepsilon_t [Z]}\bigr)\Bigr)^{-1/2}
\end{equation}
is determined by the dispersion of the wave packets. This interpretation 
follows from the fact that setting $\delta\bm{p}_t$ and $\delta\bm{q}_t$ to 
zero in (\ref{fidelitycalc2}), and therefore removing the phase space 
translations that arise from (\ref{scltimeevol}), the integral would exactly 
yield (\ref{F_disp}). In fact, $F_{\mathrm{disp}}(t)$ measures the differences 
in the dispersions caused by the two dynamics in question. This contribution 
is independent of $\hbar$. The time dependence of (\ref{F_disp}) follows from 
the relation (\ref{Gtrans}) with $S^0_t$ and $S^\varepsilon_t$, respectively. 
It is therefore completely determined by the linear stabilities of the 
perturbed and the unperturbed classical trajectory. In addition to this, the 
second factor $F_{\mathrm{class}}(t)$ is influenced by the actual separation 
$(\delta\bm{p}_t,\delta\bm{q}_t)$ of these trajectories. It reads
\begin{equation}
\label{F_class}
 F_{\mathrm{class}}(t) = 
 2^d\,\exp\Bigl[ -\frac{1}{\hbar}(\delta\bm{p}_t,\delta\bm{q}_t)\cdot 
 G_{S^0_t[Z]}\bigl(\openone - \Gamma_{t,\varepsilon}^{-1}\bigr)
 (\delta\bm{p}_t,\delta\bm{q}_t)\Bigr] \ ,\qquad 
 \Gamma_{t,\varepsilon} = \openone +G_{S^0_t [Z]}^{-1}G_{S^\varepsilon_t [Z]}
\end{equation}
and, despite of its $\hbar$-dependence, essentially represents a classical
fidelity since it is localised on the separation of the classical
trajectories: If one divides $F(t)$ by $(2\pi\hbar)^d$ as discussed
above, the contributions of (\ref{F_disp}) and (\ref{F_class}) converge to
$\delta (\delta\bm{p}_t,\delta\bm{q}_t)$ as $\hbar\to 0$. This also explains
the necessity of $\hbar$ in (\ref{F_class}). We remark that expressions
equivalent to (\ref{factor})-(\ref{F_class}) have independently been obtained 
by M.~Combescure and D.~Robert \cite{ComRob05a}.

At this point we stress that in general the separation 
$(\delta\bm{p}_t,\delta\bm{q}_t)$ of the trajectories for large $t$ differs 
essentially from the corresponding behaviour of the linearised dynamics.
In particular, an exponential instability expressed in terms of positive 
Lyapunov exponents does not imply an exponential growth of the norm of
$(\delta\bm{p}_t,\delta\bm{q}_t)$. In fact, for the dynamics of a bound
system this quantity obviously is bounded. But even then the exponent in
(\ref{F_class}) will often grow exponentially due to the presence of
the stability matrices $S^0_t$ and $S^\varepsilon_t$. 

The contributions of $F_{\mathrm{disp}}(t)$ and of $F_{\mathrm{class}}(t)$
to the decay of fidelity will now be studied separately. This procedure 
makes sense if $\hbar$ simultaneously approaches zero in order to maintain 
the condition $t\ll T_E(\hbar)$. In this regime the individual contributions 
to $F(t)$ determine the leading behaviour of the fidelity as $t\to\infty$
and $\hbar\to 0$ completely.
\subsection{Contribution of wave packet dispersion}
\label{sec2a}
We begin with discussing the behaviour of $F_{\mathrm{disp}}(t)$ as 
$t\to\infty$. Since both $G_{S^0_t [Z]}$ and $G_{S^{\varepsilon}_t [Z]}$ 
are symmetric and positive-definite, we can convert these matrices into a 
diagonal form as explained in Appendix~\ref{appb}. This implies that there 
exists a real, invertible matrix $M_t$ such that 
$M_t^T G_{S^0_t [Z]} M_t =\openone$, and at the same time 
$M_t^T G_{S^\varepsilon_t [Z]} M_t =D_t$ is diagonal, with the (positive) 
eigenvalues $\Lambda_k(t)$ of
\begin{equation}
\label{Gprod}
 G_{S^0_t [Z]}^{-1}G_{S^\varepsilon_t [Z]}=S_t^0 G_Z^{-1}
 ((S^\varepsilon_t)^{-1}S^0_t)^T G_Z (S^\varepsilon_t)^{-1}
\end{equation}
on the diagonal. Furthermore, since $G_Z$ is symmetric and positive-definite, 
it is a square of a symmetric and positive-definite matrix, $G_Z=\gamma^2$. 
Hence, (\ref{Gprod}) is conjugate to a matrix $N_t^T N_t$, with 
$N_t =\gamma\,(S^\varepsilon_t)^{-1}S^0_t\,\gamma^{-1}$. This means that 
the eigenvalues $\Lambda_k(t)$ of (\ref{Gprod}) are squares of the singular 
values $\mu_k(t)$ of $N_t$ (see Appendix~\ref{appc}). Since $\gamma$ is 
independent of $t$ the time dependence of $\mu_k(t)$ therefore is 
asymptotically determined by the singular values $\tilde\mu_k(t)$ of 
$(S^\varepsilon_t)^{-1}S^0_t$. More precisely, there exist constants
$C_1,C_2>0$ such that
\begin{equation}
\label{svasymptot}
 C_1\tilde\mu_k(t) \leq \mu_k(t) \leq C_2\tilde\mu_k(t) \ .
\end{equation}
We also exploit the fact that the product $(S^\varepsilon_t)^{-1}S^0_t$ of 
two symplectic matrices is again symplectic. This implies that its singular 
values $\tilde\mu_k(t)$ arise in pairs of mutually inverse numbers. Thus, 
they can be ordered as in (\ref{singordersymp}).

The determinant that yields $F_{\mathrm{disp}}(t)$ according to (\ref{F_disp})
can be evaluated as in Appendix~\ref{appb}, see (\ref{simdiagdet}). Taking 
into account that $G_{S^0_t [Z]}$ has unit determinant and that the 
eigenvalues $\Lambda_k(t)$ of (\ref{Gprod}) are given by squares of the 
singular values $\mu_k(t)$, we obtain
\begin{equation}
\label{F_dispcalc}
 F_{\mathrm{disp}}(t) = \left[ \prod_{k=1}^{2d}\bigl( 1+
 \mu_k(t)^2 \bigr)\right]^{-1/2} \ .
\end{equation}
The estimates (\ref{svasymptot}) then allow to bound (\ref{F_dispcalc})
from below and above in terms of
\begin{equation}
\label{F_dispbound}
 \left[ \prod_{k=1}^{d}\bigl( \tilde\mu_k(t)^2 +2 + \tilde\mu_k(t)^{-2} 
 \bigr)\right]^{-1/2} \ .
\end{equation}
More specifically, there exist constants $C_3,C_4>0$ such that
\begin{equation}
\label{F_dispest1}
 C_3\,\prod_{k=1}^{d}\tilde\mu_k(t)^{-1} \leq F_{\mathrm{disp}}(t) 
 \leq C_4\,\prod_{k=1}^{d}\tilde\mu_k(t)^{-1} \ .
\end{equation}
Since the product over the inverse singular values contains only factors
with $\tilde\mu_k(t)\geq 1$, one can introduce the simple estimate
\begin{equation}
\label{F_dispest1a}
 \tilde\mu_{\mathrm{max}}(t) \leq \prod_{k=1}^{d}\tilde\mu_k(t) \leq
 \tilde\mu_{\mathrm{max}}(t)^{d-1}\,\tilde\mu_d (t) \ .
\end{equation}
The quantities $\tilde\mu_k(t)$ are singular values of a product of two 
symplectic matrices to which the inequalities (\ref{inequal}) may be applied. 
Thus, when choosing $k=1$ in (\ref{inequal}), the l.h.s. of 
(\ref{F_dispest1a}) can be bounded from below by 
\begin{equation}
\label{F_dispest1b}
 \max \left\{ \frac{\mu_{\mathrm{max}}(S_t^0)}{\mu_{\mathrm{max}}
 (S_t^\varepsilon)},\frac{\mu_{\mathrm{max}}(S_t^\varepsilon)}
 {\mu_{\mathrm{max}}(S_t^0)}\right\} \leq 
 \tilde\mu_{\mathrm{max}}(t) \ .
\end{equation}
Furthermore, the maximal singular values of $S^\varepsilon_t$ and $S^0_t$ 
determine the maximal Lyapunov exponents according to 
\begin{equation}
\label{maxLyap}
\begin{split}
 \lambda^{0/\varepsilon} 
 &= \limsup_{t\to\infty}\frac{1}{t}\log\|S^{0/\varepsilon}_t\|_{\mathrm{HS}}\\
 &= \limsup_{t\to\infty}\frac{1}{t}\log\mu_{\mathrm{max}}
    (S^{0/\varepsilon}_t) \ ,
\end{split}
\end{equation}
so that in case $\delta\lambda = \lambda^{\varepsilon} - \lambda^0\neq 0$ 
the l.h.s. of (\ref{F_dispest1b}) is asymptotic to $e^{|\delta\lambda|t}$ 
as $t\to\infty$.

The r.h.s. of (\ref{F_dispest1a}) may now be estimated in a similar manner: 
Apply the rightmost inequality in (\ref{inequal}) to each factor, and for 
the term with $k=d$ use that $\mu_d(S_t^{0/\varepsilon})=1$, see 
Appendix~\ref{appc}. This finally yields the upper bound
\begin{equation}
\label{upbound}
 \bigl[ \mu_{\mathrm{max}}(S_t^0)\,\mu_{\mathrm{max}}(S_t^\varepsilon)
 \bigr]^{d-1}\cdot\min\bigl\{ \mu_{\mathrm{max}}(S_t^0),\mu_{\mathrm{max}}
 (S_t^\varepsilon) \bigr\} 
\end{equation}
for $\tilde\mu_{\mathrm{max}}(t)^{d-1}\tilde\mu_d(t)$. Asymptotically, as 
$t\to\infty$ this approaches 
$\exp\{[(d-1)(\lambda^0 +\lambda^\varepsilon) +\min\{\lambda^0,
\lambda^\varepsilon\}]t\}$.

The above estimates can be summarised to provide the following statement
about the asymptotic behaviour of $F_{\mathrm{disp}}(t)$: There exist 
constants $C_5,C_6>0$ such that
\begin{equation}
\label{F_dispest3}
 C_5\,e^{-Lt} \leq F_{\mathrm{disp}}(t) \leq C_6\,e^{-Lt} \ ,
\end{equation}
with 
\begin{equation}
\label{deltaLyap}
 |\delta\lambda|\leq L\leq (d-1)(\lambda^0 +\lambda^\varepsilon) 
 +\min\{\lambda^0,\lambda^\varepsilon\} \ . 
\end{equation}
Thus, once the maximal Lyapunov 
exponent of the perturbed dynamics differs from the unperturbed one, the 
asymptotic decay of $F_{\mathrm{disp}}(t)$ is essentially exponential. 
\subsection{Contribution of classical trajectories}
\label{sec2b}
The remaining factor $F_{\mathrm{class}}(t)$ that determines the decay of
fidelity is influenced by the linear stabilities of the perturbed and of 
the unperturbed classical trajectories as well as by the separation 
$(\delta\bm{p}_t,\delta\bm{q}_t)$ of the trajectories. The contribution of 
the stabilities can be treated in a similar manner as above, whereas 
the behaviour of the separation is largely unknown in a general linearly 
unstable system. Precise estimates are rare, but can possibly be achieved 
in particular cases (see, e.g., Section~\ref{sec3} and \cite{Com05}). 

A first simplification of the expression (\ref{F_class}) can be achieved
by introducing the matrix $M_t$ that converts $G_{S^0_t [Z]}$ and 
$G_{S^{\varepsilon}_t [Z]}$ into a diagonal form. The exponent of 
(\ref{F_class}), without the factor $-1/\hbar$, then reads
\begin{equation}
\label{expdiag}
 M_t^{-1}(\delta\bm{p}_t,\delta\bm{q}_t)\cdot \bigl( \openone-(\openone +
 D_t)^{-1}\bigr)M_t^{-1}(\delta\bm{p}_t,\delta\bm{q}_t) 
\end{equation}
where, as above, $D_t =M_t^T G_{S^\varepsilon_t [Z]} M_t$ is the diagonal 
matrix with the eigenvalues $\Lambda_k(t)> 0$ on its diagonal. The quadratic 
form $\openone-(\openone + D_t)^{-1}$  defined by (\ref{expdiag}) is 
positive-definite; its eigenvalues are $\Lambda_k(t)/(1+\Lambda_k(t))>0$. 
Thus, whatever value the separation $(\delta\bm{p}_t,\delta\bm{q}_t)$ attains, 
one immediately concludes that $F_{\mathrm{class}}(t)\leq 2^d$. And although 
the eigenvalues $\Lambda_k(t)$ may eventually grow as $t\to\infty$, this 
quadratic form remains bounded. Any influence of the linear stabilities on 
$F_{\mathrm{class}}(t)$ hence is encoded in the matrices $M_t$ as they appear 
in (\ref{expdiag}).

An upper bound for the expression (\ref{expdiag}) follows from the estimate
(\ref{upbound1}) derived in Appendix~\ref{appc} when choosing 
$A=\openone-(\openone +D_t)^{-1}$ and $B=M_t^{-1}$. For the calculation of 
$\|A\|_{\mathrm{tr}}$ we notice that this is given by the sum of the 
eigenvalues $\Lambda_k(t)/(1+\Lambda_k(t))$. These can be grouped in pairs
with $\Lambda_k(t)$ and  $\Lambda_k(t)^{-1}$ since the latter are eigenvalues 
of a symplectic matrix. Thus
\begin{equation}
\label{Atr}
 \|A\|_{\mathrm{tr}} = \sum_{k=1}^d \left( \frac{\Lambda_k(t)}{1+\Lambda_k(t)}
 + \frac{\Lambda_k(t)^{-1}}{1+\Lambda_k(t)^{-1}} \right) = d \ .
\end{equation}
Furthermore, we observe that
\begin{equation}
\label{BHS1}
 \| M_t^{-1} \|_{\mathrm{HS}}^2 = \tr \bigl(M_t M_t^T\bigr)^{-1} =
 \| G_{S^0_t [Z]} \|_{\mathrm{tr}} \ .
\end{equation}
Using (\ref{Gtrans}), the rightmost expression can be factorised with the 
help of (\ref{normest}), leading to  
\begin{equation}
\label{BHS2}
 \| M_t^{-1} \|_{\mathrm{HS}}^2 
 \leq \| S_t^0 \|_{\mathrm{HS}}^2 \,\|G_Z\|_{\mathrm{HS}}\ .
\end{equation}
Our final upper bound for (\ref{expdiag}) therefore reads
\begin{equation}
\label{upbdexp}
 d\,\|G_Z\|_{\mathrm{HS}}\,\| S_t^0 \|_{\mathrm{HS}}^2 \,
 \bigl( \delta\bm{p}_t^2 + \delta\bm{q}_t^2 \bigr) \ .
\end{equation}
For the contribution (\ref{F_class}) to the fidelity this provides us with
a lower bound of the form
\begin{equation}
\label{Fclasslow}
 F_{\mathrm{class}}(t) \geq 2^d\,\exp\Bigl[ -\frac{d}{\hbar}
 \|G_Z\|_{\mathrm{HS}}\,\| S_t^0 \|_{\mathrm{HS}}^2 \,
 \bigl( \delta\bm{p}_t^2 + \delta\bm{q}_t^2 \bigr) \Bigr] \ .
\end{equation}
In addition, if the unperturbed trajectory possesses a positive maximal 
Lyapunov exponent, the factor $\| S_t^0 \|_{\mathrm{HS}}^2$ grows 
asymptotically like $e^{2\lambda^0 t}$, see (\ref{maxLyap}).

In order to achieve a lower bound for (\ref{expdiag}) according to 
(\ref{lowbound}) we first notice that the symplecticity of (\ref{Gprod}) 
implies $\Lambda_{\mathrm{min}}(t)=\Lambda_{\mathrm{max}}(t)^{-1}=
\mu_{\mathrm{max}}(t)^{-2}$. This leads to 
\begin{equation}
\label{lambdamin1}
 \Lambda_{\mathrm{min}}(A) = \frac{1}{1+\mu_{\mathrm{max}}(t)^2} 
 \geq \frac{1}{2}\,\mu_{\mathrm{max}}(t)^{-2}  \ .
\end{equation}
Then (\ref{svasymptot}) and (\ref{Fan}) yield the further bound  
\begin{equation}
\label{lambdamin2}
\begin{split}
 \Lambda_{\mathrm{min}}(A) 
 &\geq K_1\,\mu_{\mathrm{max}}\bigl( (S^\varepsilon_t)^{-1}S^0_t\bigr)^{-2} \\
 &\geq K_1\,\mu_{\mathrm{max}}(S^\varepsilon_t)^{-2}\,
       \mu_{\mathrm{max}}(S^0_t)^{-2} \ 
\end{split}
\end{equation}
with some constant $K_1>0$. We now require a lower bound for
$\mu_{2d}(B)^2 =\mu_{2d}(M_t^{-1})^2$, and first notice that this quantity is 
the lowest eigenvalue of $(M_t M_t^T)^{-1}=G_{S^0_t[Z]}$, which in turn is 
the inverse of the largest eigenvalue of this matrix. Making use of the 
relation (\ref{Fan}) we then conclude that
\begin{equation}
\label{muminB}
\begin{split}
 \Lambda_{\mathrm{max}}\bigl( G_{S^0_t[Z]}\bigr)
 &=    \mu_{\mathrm{max}}\bigl( ((S^0_t)^{-1})^T G_Z (S^0_t)^{-1}\bigr) \\
 &\leq \mu_{\mathrm{max}}(G_Z) \, \mu_{\mathrm{max}} (S^0_t)^2 \ .
\end{split}
\end{equation}
Collecting the above estimates therefore provides us with the lower bound
\begin{equation}
\label{lowbdexp}
 K_2\,\mu_{\mathrm{max}}(S^\varepsilon_t)^{-2}\,\mu_{\mathrm{max}}(S^0_t)^{-4}
 \bigl( \delta\bm{p}_t^2 + \delta\bm{q}_t^2 \bigr)
\end{equation}
for (\ref{expdiag}), with some $K_2 >0$. Hence,
\begin{equation}
\label{Fclassup}
 F_{\mathrm{class}}(t) \leq 2^d\,\exp\Bigl[ -\frac{K_2}{\hbar}
 \frac{\delta\bm{p}_t^2 + \delta\bm{q}_t^2}{\mu_{\mathrm{max}}
 (S^\varepsilon_t)^{2}\,\mu_{\mathrm{max}}(S^0_t)^{4}} \Bigr] \ .
\end{equation}
Furthermore, in the case of linearly unstable trajectories (\ref{maxLyap}) 
implies that
\begin{equation}
\label{lowbdLyap}
 \mu_{\mathrm{max}}(S^\varepsilon_t)^{2}\,\mu_{\mathrm{max}}(S^0_t)^{4}
 \sim \exp [(2\lambda^\varepsilon +4\lambda^0)t] \ ,
\end{equation}
as $t\to\infty$.

We have so far refrained from estimating the squared distance 
$\delta\bm{p}_t^2 +\delta\bm{q}_t^2$ the perturbed trajectory can separate 
itself away from the unperturbed one. However, any further statements
about the decay of $F_{\mathrm{class}}(t)$ require some knowledge of the
behaviour of that distance. But this, as already mentioned, seems to be 
difficult to obtain. E.g., one can in general not exclude that this quantity 
vanishes infinitely often (see \cite{Com05} for a particular case), or
asymptotically approaches zero. Such situations may occur, if the 
perturbation $V(\bm{p},\bm{q})$ is confined to a bounded part of phase space, 
but the trajectories are forced to leave this domain in a particular channel.  
According to (\ref{F_class}), at those instances where  
$\delta\bm{p}_t^2 +\delta\bm{q}_t^2$ vanishes, $F_{\mathrm{class}}(t)$ clearly 
acquires its maximal possible value $2^d$. On the other hand, if the energy 
shells corresponding to both the perturbed and the unperturbed classical 
Hamiltonians are bounded, $\delta\bm{p}_t^2 +\delta\bm{q}_t^2$ will 
necessarily be bounded, too. 

However, if the classical motion is unbounded (as in the example in 
Section~\ref{sec3}), an upper bound for this distance would be helpful. In 
order to achieve such an estimate we consider a Taylor expansion of 
$(\bm{p}_t^\varepsilon,\bm{q}_t^\varepsilon)$ about $\varepsilon =0$ with 
remainder term of first order, i.e.,
\begin{equation}
\label{Taylor1}
 (\delta\bm{p}_t,\delta\bm{q}_t) = -\theta\varepsilon\,\left.
 \frac{d}{d\varepsilon'}(\bm{p}_t^{\varepsilon'},\bm{q}_t^{\varepsilon'})
 \right|_{\varepsilon' =\theta\varepsilon} \ ,
\end{equation}
where $\theta\in [0,1]$. The derivative on the r.h.s. can now be identified 
as a solution of a differential equation in the variable $t$. Abbreviating
\begin{equation}
\label{zdef}
 \bm{z}_\varepsilon(t) = 
 \frac{d}{d\varepsilon}(\bm{p}_t^{\varepsilon},\bm{q}_t^{\varepsilon}) \ ,
\end{equation}
the fact that $(\bm{p}^\varepsilon_0,\bm{q}^\varepsilon_0) = 
(\bm{p},\bm{q})$ for all $\varepsilon$ implies the initial condition 
$\bm{z}_\varepsilon(0)=0$. Moreover, a derivative of Hamilton's equations
of motion
\begin{equation}
\label{Hameq}
 \bigl(\dot{\bm{p}}^\varepsilon_t,\dot{\bm{q}}^\varepsilon_t\bigr) =
 J\,H'_\varepsilon \bigl(\bm{p}^\varepsilon_t,\bm{q}^\varepsilon_t\bigr)
\end{equation}
w.r.t. $\varepsilon$ yields
\begin{equation}
\label{deq4z}
 \dot{\bm{z}}_\varepsilon(t) = 
 J\,H''_\varepsilon \bigl(\bm{p}^\varepsilon_t,\bm{q}^\varepsilon_t\bigr)
 \,\bm{z}_\varepsilon(t) + J\,V'\bigl(\bm{p}^\varepsilon_t,
 \bm{q}^\varepsilon_t\bigr) \ ,
\end{equation}
where $V'(\bm{p},\bm{q})$ denotes the gradient of the function 
$V(\bm{p},\bm{q})$ whose Weyl quantisation yields the perturbation $\hat V$ 
of the quantum Hamiltonian. A solution of the inhomogeneous differential 
equation (\ref{deq4z}) with the prescribed initial condition is then provided 
by the integral
\begin{equation}
\label{deltaint}
 \bm{z}_\varepsilon(t) =
 S_t^\varepsilon \int_0^t (S_s^\varepsilon)^{-1}\,J\,
 V'(\bm{p}_s^{\varepsilon},\bm{q}_s^{\varepsilon})\ ds \ .
\end{equation}
Used on the r.h.s. of (\ref{Taylor1}) this expression allows us to relate
the separation $(\delta\bm{p}_t,\delta\bm{q}_t)$ of the trajectories to
their linear stabilities and properties of the derivative of the classical
perturbation $V$.

A quantitative upper bound that immediately follows from (\ref{deltaint})
is
\begin{equation}
\label{deltabound}
\begin{split}
 0\leq |(\delta\bm{p}_t,\delta\bm{q}_t)| 
  &\leq \varepsilon\theta t\,\|S_t^{\theta\varepsilon}\|_{\mathrm{HS}}\,
        \sup_{s\in [0,t]}\|S_s^{\theta\varepsilon}\|_{\mathrm{HS}}\, 
        \frac{1}{t}\int_0^t |V'(\bm{p}_s^{\theta\varepsilon},
        \bm{q}_s^{\theta\varepsilon})|\ ds \\
  &\leq \varepsilon t\,\bigl(\Sigma_t^\varepsilon\bigr)^2 \, 
        \sup_{\theta\in [0,1]}\frac{1}{t}\int_0^t 
        |V'(\bm{p}_s^{\theta\varepsilon},\bm{q}_s^{\theta\varepsilon})|\ ds \ .
\end{split}
\end{equation}
Here we have introduced
\begin{equation}
\label{Sigma}
 \Sigma_t^\varepsilon =\sup_{s\in [0,t],\theta\in [0,1]}
  \|S_s^{\theta\varepsilon}\|_{\mathrm{HS}} \ ,
\end{equation}
whose asymptotic behaviour in the case of linearly unstable trajectories
is controlled by the exponent
\begin{equation}
\label{labar}
 \overline{\lambda}^\varepsilon = \sup_{\theta\in [0,1]}
 \lambda^{\theta\varepsilon} \ .
\end{equation}
Furthermore, under favourable circumstances the time average 
$\overline{V'}$ of $V'$ in (\ref{deltabound}) is finite as $t\to\infty$;
then the asymptotic behaviour of the r.h.s. in (\ref{deltabound}) for
large times is given by
\begin{equation}
\label{delasympt}
 \varepsilon\,\overline{V'}\,t\,e^{2\overline{\lambda}^\varepsilon t}\ .
\end{equation}
This will, e.g., be the case if either the trajectory remains in a bounded 
set, or the derivative $V'$ is a bounded function on the respective energy 
shell. 
   
\section{Inverted oscillators}
\label{sec3}
We want to discuss a simple and exactly solvable example that nevertheless 
possesses the typical features of exponentially unstable classical dynamics: 
a $d$-dimensional inverted harmonic oscillator. In that case the Hamiltonian
is quadratic in position and momentum and therefore the semiclassical 
propagation (\ref{scltimeevol}) is exact, i.e., the error term vanishes. 

To be specific, let
\begin{equation}
\label{invoscillator}
 H_0(\bm{p},\bm{q}) = \frac{1}{2}\bm{p}^2 - \frac{\omega^2}{2}\bm{q}^2 \ ,
\end{equation}
and define 
\begin{equation}
\label{invoscper}
 H_\varepsilon (\bm{p},\bm{q}) = H_0(\bm{p},\bm{q}-\varepsilon\bm{a}) \ , 
\end{equation}
so that, up to a constant, $V(\bm{q})=\omega^2\bm{a}\cdot\bm{q}$. This 
perturbation consists of a phase space translation of the unperturbed 
Hamiltonian and hence is of the same type as the one discussed in 
\cite{BevHel04}. The equations of motion generated by the unperturbed and by 
the perturbed Hamiltonian, respectively, can be solved explicitly, leading to 
\begin{equation}
\label{classtrajectory}
\begin{split}
 \bm{p}^0_t &= \bm{p}\,\cosh\omega t + \bm{q}\,\omega\sinh\omega t \ ,\\
 \bm{q}^0_t &= \bm{q}\,\cosh\omega t + \bm{p}\,\omega^{-1}\sinh\omega t \ ,
\end{split}
\end{equation}
and 
\begin{equation}
\label{deltapq}
\begin{split}
 \delta\bm{p}_t &= \bm{a}\,\varepsilon\omega\sinh\omega t \ ,\\
 \delta\bm{q}_t &= \bm{a}\,\varepsilon (\cosh\omega t -1) \ .
\end{split}
\end{equation}
From (\ref{Salt}), (\ref{classtrajectory}), and (\ref{deltapq}) one, 
moreover, obtains 
\begin{equation}
\label{Sinvosc}
 S^\varepsilon_t = S^0_t =
 \begin{pmatrix} \cosh\omega t\,\openone & \omega\sinh\omega t\,\openone \\
 \omega^{-1}\sinh\omega t\,\openone & \cosh\omega t\,\openone\end{pmatrix}\ .
\end{equation}
This implies an accidental coincidence of the unperturbed and the perturbed 
Lyapunov exponents: $\lambda^0 = \omega = \lambda^\varepsilon$. Furthermore,
$G_{S^0_t [Z]}=G_{S^\varepsilon_t [Z]}$ so that $F_{\mathrm{disp}}(t)=2^{-d}$,
reflecting the fact that the coinciding perturbed and unperturbed linearised 
dynamics lead to the same dispersions of the wave packets. Since
$\delta\lambda =0$, this finding is in accordance with the bounds 
(\ref{deltaLyap}). With the help of the relation (\ref{Wignermeta}) and a 
change of variables the expression (\ref{fidelitycalc2}) for the fidelity can 
now be brought into the form
\begin{equation}
\label{fidcalcinvosc}
 F(t) = \iint W[\phi^\hbar]\bigl((\bm{\eta},\bm{y})-S_t^{-1} 
 (\delta\bm{p}_t,\delta\bm{q}_t)\bigr)\,W[\phi^\hbar](\bm{\eta},\bm{y})\ 
 \frac{d\eta\,dy}{(2\pi\hbar)^d} \ .
\end{equation}
Therefore, in this example the quantum fidelity is crucially determined 
by the separation (\ref{deltapq}) of the classical trajectories. 

For simplicity one can imagine the initial wave packet to be a Gaussian 
(\ref{Gauss}) with $Z=i\openone$ that is localised at the unstable fixed 
point $(\bm{p},\bm{q})=(0,0)$ of the unperturbed classical dynamics. Thus, 
$(\bm{p}^0_t,\bm{q}^0_t)=(0,0)$ for all $t$, so that the unperturbed 
time evolution (\ref{scltimeevol}) fixes the centre of the wave packet and 
only forces it to disperse according to the action of the metaplectic operator 
related to (\ref{Sinvosc}). The perturbed dynamics, however, pushes the 
centre away from the fixed point according to (\ref{deltapq}). This happens 
with an exponential rate that follows from the asymptotic behaviour 
\begin{equation}
\label{distasymp}
 |(\delta\bm{p}_t,\delta\bm{q}_t)| \sim \frac{\varepsilon |\bm{a}|}{2}\,
 \sqrt{\lambda^2 +1}\ e^{\lambda t}\ , \quad t\to\infty\ ,
\end{equation}
of the distance, which may be compared with the corresponding 
asymptotics
\begin{equation}
\label{delasymptinosc}
 \varepsilon |\bm{a}|\lambda^2\,t\,e^{2\lambda t}
\end{equation}
of the upper bound (\ref{deltabound}), see also (\ref{delasympt}). 

For Gaussian states either (\ref{F_class}) or (\ref{fidcalcinvosc}) can be
evaluated directly, yielding
\begin{equation}
\label{Finvosc}
 F(t) = \exp\Bigl[ -\frac{\varepsilon^2\bm{a}^2}{2\hbar}\bigl(
 (1-\cosh\lambda t)^2-\lambda^2\sinh^2\lambda t \bigr)\Bigr] \ .
\end{equation}
In this example the quantum fidelity therefore decays extremely fast, namely 
in a double exponential manner. We stress that neither have approximations 
been performed nor have any assumptions entered, and hence the result holds 
unconditionally. Clearly, this finding is at variance with the previous 
predictions of an exponential decay of fidelity. However, it obviously
complies with the bounds (\ref{Fclasslow}) and (\ref{Fclassup}) derived 
in Section~\ref{sec2}. We remark that the double exponential decay is caused 
by both the separation (\ref{deltapq}) of the trajectories and the exponential 
instability of the linearised dynamics (\ref{Sinvosc}). Each of these factors 
alone would lead to this effect.

\section{Conclusions}
\label{sec4}
Our approach to the quantum fidelity of localised wave packets led us to
distinguish two effects that derive from two separate contributions to the
semiclassical evolution prior to the Ehrenfest time. 

One effect is caused by the dispersion of the wave packets, which 
semiclassically originates from the metaplectic representation of the 
linearised classical dynamics. Since generically the unperturbed and the 
perturbed classical dynamics possess different linearisations, the resulting 
non-coinciding dispersions cause an eventually exponential contribution 
$F_{\mathrm{disp}}(t)$ to the decay of fidelity as described by 
(\ref{F_dispest3}) and (\ref{deltaLyap}).
 
A second effect is due to the separation of the perturbed classical 
trajectory away from the unperturbed one. Since the centres of the wave
packets follow their associated classical trajectories, this divergence
forces the overlap of the unperturbed with the perturbed time evolution
of the given initial state to decrease. We estimated this contribution,
$F_{\mathrm{class}}(t)$, for Gaussian wave packets and identified an 
influence of the linear stabilities as well as of the separation of the 
trajectories. Since in general the latter cannot be well controlled, we 
were unable to determine a uniform expression for this factor. The bounds
(\ref{Fclasslow}), (\ref{Fclassup}), and (\ref{deltabound}) that we obtained 
allow for decays that are exponentially faster, or slower, than exponential. 
Of course, the further factor $F_{\mathrm{disp}}(t)$ always ensures that
the fidelity decays at least exponentially.

In view of the previous predictions of an exponential fidelity decay a 
contribution that decreases in a double exponential manner might come as a 
surprise. In the example of inverted harmonic oscillators, however, we
saw that such a behaviour is indeed possible. In that case this was caused
by both the linear instability of the classical motion and by the 
exponentially growing separation of the trajectories. The latter effect
is certainly not generic if one has chaotic systems with bounded energy
shells in mind. Nevertheless, in our example the linear instability alone 
would cause a doubly exponentially decreasing factor. And this is in perfect 
agreement with the bound (\ref{Fclasslow}) that applies in the general
case, even if the separation of trajectories does not exceed a given bound.

\subsection*{Acknowledgments}
We would like to thank Rainer Glaser for many useful discussions and
Monique Combescure for communicating \cite{ComRob05b,ComRob05a} prior to
publication.

\vspace*{0.5cm}

\begin{appendix}
\section*{Appendices}
\section{Metaplectic representation}
\label{appa}
In this appendix we collect some important facts about the symplectic group
and the metaplectic representation. For further details see 
\cite{Lit86,Fol89,ComRob05b}.

The \textit{symplectic group} consists of the linear canonical transformations
$(\bm{p},\bm{q})\mapsto(\bm{p}',\bm{q}')$ with
\begin{equation}
\label{lincanon}
\begin{split}
 \bm{p}' &= A\bm{p} + B\bm{q} \ , \\
 \bm{q}' &= C\bm{p} + D\bm{q} \ .
\end{split}
\end{equation}
The real $2d\times 2d$ matrix
\begin{equation}
\label{sympmatrix}
 S =\begin{pmatrix}A&B\\C&D\end{pmatrix}
\end{equation}
then fulfills $S^T JS=J$, where 
\begin{equation}
\label{Jdef}
 J=\begin{pmatrix}0&-\openone\\ \openone&0\end{pmatrix} \ , 
 \qquad J^2=-\openone \ ,
\end{equation}
is the symplectic unit. The symplectic group is generated by the matrices
\begin{equation}
\label{sympgen}
 S_A =\begin{pmatrix} A & 0 \\ 0 & (A^T)^{-1} \end{pmatrix} \ ,\quad
 S_C =\begin{pmatrix} \openone & C \\ 0 & \openone \end{pmatrix} \ ,
      \quad J \ , 
\end{equation}
where $A$ is an invertible matrix and $C$ is symmetric.

A quantisation of a linear canonical transformation (\ref{lincanon}) requires
a unitary ray-re\-presentation of the symplectic group. This can be obtained
from the observation that the operators $\hat D (\bm{p},\bm{q})$ and 
$\hat D( (S^T)^{-1}(\bm{p},\bm{q}))$, see (\ref{phasetrans}), each provide a 
unitary irreducible representation of the Heisenberg group. According to the 
Stone-Von~Neumann~Theorem there hence exists a unitary operator $\hat\mu(S)$ 
such that
\begin{equation}
\label{metadef}
 \hat D\bigl( (S^T)^{-1}(\bm{p},\bm{q}) \bigr) = 
 \hat\mu(S)\,\hat D (\bm{p},\bm{q})\,\hat\mu(S)^{-1} \ .
\end{equation}
Choosing $S=S_1 S_2$ furthermore implies the multiplicative property
\begin{equation}
\label{metamult}
 \hat\mu(S_1 S_2) = e^{i\chi(S_1 ,S_2)}\,\hat\mu(S_1)\,\hat\mu(S_2)\ .
\end{equation}
In fact, the phase factor can be chosen to be $\pm 1$. The 
\textit{metaplectic operators} $\hat\mu(S)$ determine a double-valued 
representation of the symplectic group which is also known as the 
\textit{metaplectic representation}. 

Up to a sign the metaplectic representation is fixed once the metaplectic 
operators for the generators (\ref{sympgen}) are given. Exploiting the
relation (\ref{metadef}), one obtains
\begin{equation}
\label{metagen}
\begin{split}
 \hat\mu(S_A)\psi (\bm{x}) &= \sqrt{\det A}\,\psi(A^T\bm{x}) \ , \\
 \hat\mu(S_C)\psi (\bm{x}) &= \pm e^{\frac{i}{2\hbar}\bm{x}\cdot C\bm{x}}\,
                              \psi(\bm{x})\ , \\
 \hat\mu(J)\psi (\bm{p})   &= i^{d/2}\,\widehat\psi(\bm{p})\ , 
\end{split}
\end{equation}
for them, where $\widehat\psi(\bm{p})$ denotes the momentum representation 
of $\psi$.

An explicit calculation based on the relation (\ref{metadef}) finally
reveals that the Wigner representation of a quantum state is covariant under
linear canonical transformations,
\begin{equation}
\label{Wignermeta}
 W[\hat\mu(S)\psi](\bm{\xi},\bm{x}) = 
 W[\psi]\bigl( S^{-1}(\bm{\xi},\bm{x}) \bigr) \ .
\end{equation}
Thus, if $\psi$ is localised at the point $(\bm{p},\bm{q})$ in phase space,
the transformed state $\hat\mu(S)\psi$ is concentrated at $S(\bm{p},\bm{q})$.
\section{Diagonal form of positive matrices}
\label{appb}
It is well known that if two real and symmetric matrices commute, they can 
be simultaneously diagonalised by an orthogonal transformation. Less 
appreciated is the possibility of converting non-commuting, but 
positive-definite, symmetric matrices into a diagonal form with a single
transformation:

Let $A$ and $B$ be real, symmetric, and positive-definite $n\times n$
matrices. Then there exists a real, invertible (not necessarily orthogonal)
matrix $M$ such that $M^T AM=\openone$ and $M^T BM$ is diagonal, with 
the eigenvalues $\Lambda_j(A^{-1}B)$ of the positive-definite matrix 
$A^{-1}B$ on the diagonal. Moreover,
\begin{equation}
\label{simdiagdet}
 \det (A+B) = \prod_{j=1}^n \Lambda_j (A) \bigl[ 1+\Lambda_j(A^{-1}B)
 \bigr] \ .
\end{equation}
A proof of this statement is not difficult: Let $O$ be an orthogonal matrix
such that $O^T AO =D$ is diagonal (and positive-definite). Define 
$U = OD^{-1/2}$, then $U^T A U =\openone$, and $U^T B U$ is symmetric and 
positive-definite. Furthermore, since $U^T B U = U^{-1}A^{-1}BU$, the matrices
$U^T B U$ and $A^{-1}B$ have identical spectra. Hence there exists an 
orthogonal matrix $O_1$ such that $O_1^T U^T BUO_1$ is diagonal, with the
eigenvalues $\Lambda_j(A^{-1}B)$ on the diagonal. Then define $M=UO_1$ to
obtain the matrix $M$ of the statement.
\section{Matrix norms and singular values}
\label{appc}
Real $n\times n$ matrices can be estimated in terms of various norms,
for which there exists a number of inequalities that we want to review
in this appendix. More details can, e.g., be found in \cite{Sim79}.

The \textit{operator norm} is defined as
\begin{equation}
\label{opnorm}
 \| A \|_{\mathrm{op}} = \sup_{|\bm{x}|=1} |A\bm{x}| \ ,
\end{equation}
where $|\bm{x}|=\sqrt{\bm{x}^2}$ is the euclidean norm of a vector
$\bm{x}\in\mathbb{R}^n$. The \textit{trace norm}, however, is given by
\begin{equation}
\label{trnorm}
 \| A \|_{\mathrm{tr}} = \tr \sqrt{A^T A} \ .
\end{equation}
Finally, we consider the \textit{Hilbert-Schmidt norm}
\begin{equation}
\label{HSnorm}
 \| A \|_{\mathrm{HS}} = \sqrt{\tr A^T A} \ ,
\end{equation}
All of these matrix norms possess the multiplicative property
$\| AB \| \leq \| A\| \| B \|$. Moreover, they fulfill
\begin{equation}
\label{normest}
\begin{split}
 &\| A \|_{\mathrm{op}} \leq \| A \|_{\mathrm{HS}} \leq 
  \| A \|_{\mathrm{tr}} \ , \\
 &\| AB \|_{\mathrm{tr}} \leq \| A \|_{\mathrm{HS}} \| B \|_{\mathrm{HS}} \ .
\end{split}
\end{equation}
In addition, for symplectic matrices $S$ one obtains 
$\| S^{-1} \|_{\mathrm{tr}/\mathrm{HS}} = \| S \|_{\mathrm{tr}/\mathrm{HS}}$.

In general a real $n\times n$ matrix $A$ cannot be diagonalised. However,
$A^T A$ is symmetric and positive-definite and therefore possesses $n$
non-negative eigenvalues. Their positive square-roots $\mu_j (A)$ are the
\textit{singular values} of $A$, which we order as 
\begin{equation}
\label{singvalorder}
 \mu_{\mathrm{max}}(A) = \mu_1 (A) \geq \mu_2 (A) \geq \dots \geq
 \mu_n (A) \geq 0 \ .
\end{equation}
Furthermore, Fan's inequality (see \cite{Sim79}) implies for the singular 
values of products that 
\begin{equation}
\label{Fan}
\begin{split}
 \mu_k (AB)&\leq \mu_{\mathrm{max}}(A)\,\mu_k(B)\ , \\
 \mu_k (AB)&\leq \mu_{\mathrm{max}}(B)\,\mu_k(A)\ .
\end{split}
\end{equation}
The singular values of symplectic matrices occur in pairs of mutually 
inverse numbers. Since in that case $n$ must be even, we write $n=2d$,
and choose the following ordering:
\begin{equation}
\label{singordersymp}
 \mu_1 (S) \geq \dots \geq \mu_d (S) \geq \mu_d (S)^{-1} \geq \dots \geq 
 \mu_1 (S)^{-1} \ .
\end{equation}
In addition to the upper bound (\ref{Fan}), in the symplectic case one
can also find a lower bound that is based on the fact that 
$\mu_{\mathrm{max}}(S^{-1})=\mu_{\mathrm{max}}(S)$. Choose first $A=S_1 S_2$ 
and $B=S_2^{-1}$, and then $A=S_1^{-1}$ and $B=S_1 S_2$ in (\ref{Fan}).
This results in
\begin{equation}
\label{inequal}
 \max \left\{ \frac{\mu_k(S_1)}{\mu_{\mathrm{max}}(S_2)},\frac{\mu_k(S_2)}
 {\mu_{\mathrm{max}}(S_1)} \right\} \leq \mu_k(S_1 S_2) \leq \min\left\{
 \mu_k (S_1)\,\mu_{\mathrm{max}}(S_2),\mu_k (S_2)\,\mu_{\mathrm{max}}(S_1)
 \right\} \ .
\end{equation}
In section~\ref{sec2b} we need to estimate a quadratic form 
$B\bm{v}\cdot AB\bm{v}$ in terms of suitable norms of the positive-definite, 
symmetric matrix $A$ and of the invertible matrix $B$.
An upper bound follows from the definition (\ref{opnorm}) of the operator
norm and a subsequent application of (\ref{normest}),
\begin{equation}
\label{upbound1}
\begin{split}
 B\bm{v}\cdot AB\bm{v} 
 &\leq \|A\|_{\mathrm{op}}\,\|B\|_{\mathrm{op}}^2 |\bm{v}|^2 \\
 &\leq \|A\|_{\mathrm{tr}}\,\|B\|^2_{\mathrm{HS}}\,|\bm{v}|^2 \ . 
\end{split}
\end{equation}
A lower bound can be gained from the fact that the quadratic form defined 
by $A$ attains its minimum at the eigenvector corresponding to the lowest
eigenvalue $\Lambda_{\mathrm{min}}(A)>0$. Thus  
\begin{equation}
\label{lowbound}
\begin{split}
 B\bm{v}\cdot AB\bm{v} 
 &\geq \Lambda_{\mathrm{min}}(A)\,|B\bm{v}|^2 \\
 &=    \Lambda_{\mathrm{min}}(A)\, \bm{v}\cdot B^T B\bm{v}\\
 &\geq \Lambda_{\mathrm{min}}(A)\,\mu_n (B)^2\,|\bm{v}|^2 \ ,
\end{split}
\end{equation}
since by (\ref{singvalorder}) $\mu_n (B)$ is the lowest singular value of 
$B$.

\end{appendix}

{\small
\bibliographystyle{my_unsrt}
\bibliography{literatur}}

\end{document}